\documentclass[conference]{IEEEtran}

\usepackage{graphicx}

\usepackage{amsmath}
\usepackage{amssymb}

\usepackage{amsthm,amsfonts}
\usepackage{xcolor}
\usepackage{enumitem}
\usepackage{url}

\usepackage{enumitem}
\setlist{nosep}

\def\BibTeX{{\rm B\kern-.05em{\sc i\kern-.025em b}\kern-.08em
    T\kern-.1667em\lower.7ex\hbox{E}\kern-.125emX}}

\IEEEoverridecommandlockouts

\begin{document}

\title{To Ship or Not to (Function) Ship \\ (Extended version) \\
}

\author{
\IEEEauthorblockN{Feilong Liu, Niranjan Kamat, Spyros Blanas, Arnab Nandi}
\IEEEauthorblockA{\textit{The Ohio State University} \\
\texttt{\scriptsize\{liu.3222, kamat.14, blanas.2, nandi.9\}@osu.edu}
}
}

\maketitle

\begin{abstract}

Sampling is often used to reduce query latency for interactive big data analytics. The established parallel data processing paradigm relies on function shipping, where a coordinator dispatches queries to worker nodes and then collects the results. The commoditization of high-performance networking makes \emph{data shipping} possible, where the coordinator directly reads data in the workers' memory using RDMA while workers process other queries. In this work, we explore when to use function shipping or data shipping for interactive query processing with sampling. Whether function shipping or data shipping should be preferred depends on the amount of data transferred, the current CPU utilization, the sampling method and the number of queries executed over the data set. The results show that data shipping is up to 6.5$\times$ faster when performing clustered sampling with heavily-utilized workers.

\end{abstract}

\section{Introduction}
\label{intro}

In big data analysis, sampling is often used to reduce query latency for interactive query
execution~\cite{olken1993random}.
Current database systems use function shipping in query execution, where
the coordinator distributes query plans to the workers
for execution then collect results from the workers.
The cost of function shipping includes the computation cost of executing queries
in workers and the communication cost of transferring results from workers to the
coordinator.
In function shipping, sampling methods do not affect the communication cost, but affect
the computation cost.
For example, random sampling accesses the whole data set while cluster sampling
only accesses part of the data set during query execution.

Commodity clusters are now commonly equipped with fast networks with Remote Direct
Memory Access (RDMA) support~\cite{BarthelsRackjoin15}.
RDMA enables user applications to directly access memory in
remote machines without
involving the operating kernel and offers higher throughput than TCP/IP
sockets~\cite{FreyAicdcs09}.
Data shipping is possible with one-sided memory access provided by RDMA.
In data shipping, the coordinator uses RDMA Read to read data from workers
and executes query locally while the workers remain passive.
The cost of data shipping includes the computation cost of executing queries in the coordinator
and the communication cost of transferring data from workers to the coordinator.
In data shipping, sampling methods not only affect the computation cost, but also affect
the communication cost.
In cluster sampling only the sample of the data set is transferred to the coordinator,
however, the whole data set is transferred to the coordinator in random sampling.

In this work, we add the optimization of choosing between function shipping and data shipping
in our RDMA-aware system~\cite{liu2017design}.
We discuss the trade-offs between function shipping and data shipping that are afforded by the advent of RDMA and look at how sampling influences this decision.
Whether function shipping or data shipping should be preferred depends on the amount of data transferred, the current CPU utilization, the sampling method and the number of queries executed on the data set.
The result shows that data shipping has better performance when the computing resources are limited in
workers for both sampling methods and
data shipping improves performance by up to 6.5$\times$.

\begin{figure*}[t]
\begin{minipage}{0.31\linewidth}
\includegraphics[]{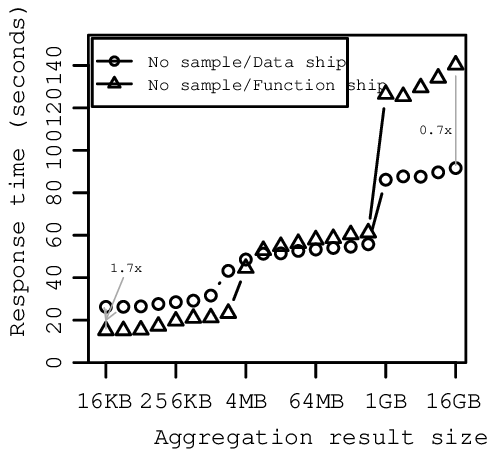}
\caption{As distinct cardinality increases, function shipping becomes expensive due to result set size increase.
}
\label{fig:varycardinality}
\end{minipage}
\hfill
\begin{minipage}{0.31\linewidth}
\includegraphics[]{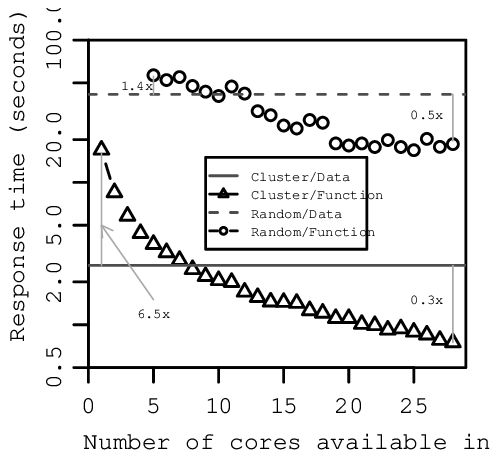}
\caption{As the computation resources available at worker increases, function shipping gets cheaper.
}
\label{fig:varycores}
\end{minipage}
\hfill
\begin{minipage}{0.31\linewidth}
\includegraphics[]{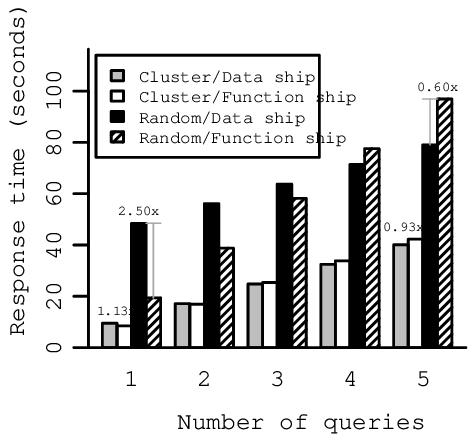}
\caption{As the result set size increases due to more queries being executed, data shipping gets cheaper.
}
\label{fig:varyquery}
\end{minipage}
\end{figure*}

\section{System Design}
\label{sys-details}

We use the traditional single coordinator, multiple workers design.
User queries are sent to the coordinator. 
The coordinator directs the query to all the workers.
Each worker can then choose to respond through either data shipping or function shipping.
In data shipping, the worker returns the raw data to the coordinator and the coordinator executes the query on the received data.
In function shipping, the worker executes the query on its data and returns the result back to the coordinator.
A final aggregation to combine the results from all workers is then performed at the coordinator. 

\subsection{Function Shipping vs Data Shipping}
\label{system:function-vs-data-shipping}

In function shipping, the worker executes the query and performs RDMA Write to send the result to the coordinator. 
In data shipping, the coordinator uses RDMA Read to read the data and executes the query on received data. The worker is passive in data shipping.
The costs of data shipping and function shipping are as follows:

\vspace{-1em}
\begin{flalign}
&\texttt{COST}(DS) = C_{Read} + C_{Sample} + C_{CExec}&
\label{eq:ds}
\end{flalign}
\vspace{-2em}
\begin{flalign}
&\texttt{COST}(FS) = C_{Sample} + C_{WExec} + C_{Write} + C_{CAgg}&
\label{eq:fs}
\end{flalign}
\vspace{-1em}

where 
\texttt{COST}$(DS)$ is the cost of data shipping, 
$C_{Read}$ is the cost of reading data from workers,
$C_{Sample}$ is the cost of sampling,
$C_{CExec}$ is the cost of executing queries at the coordinator,
\texttt{COST}$(FS)$ is the cost of function shipping, 
$C_{WExec}$ is the cost of executing queries at the worker,
$C_{Write}$ is the cost of writing the result to the coordinator,
and $C_{CAgg}$ is the cost of aggregating results from the workers.
When there are multiple workers, the workload is split and distributed across all workers.
Increasing the number of workers reduces the data to be processed in each worker, which
decreases the sampling cost $C_{Sample}$ and the execution cost at the workers $C_{WExec}$ in Equation~\ref{eq:fs},
and hence favors function shipping.

While function shipping is the norm, data shipping is preferred when the cost of function shipping
\texttt{COST}$(FS)$ is higher than the cost of data shipping \texttt{COST}$(DS)$.
According to Equation~\ref{eq:ds} and Equation~\ref{eq:fs},
$C_{Write}$ is exclusive to function shipping and depends on the size of the result.
Larger result size means higher cost for function shipping, which makes data shipping preferred.
Data shipping is also preferred when there is heavy load in workers and less computing
resources available for query execution.
The execution cost $C_{WExec}$ in Equation~\ref{eq:fs} increases when there are less computing resources
in workers, which leads to the increase of the function shipping cost \texttt{COST}$(FS)$.
Hence data shipping is preferred in the following cases:
\begin{enumerate}[label=\arabic*)]
\item The size of the result is large.
\item Computation load on the worker is high. 
\end{enumerate}

\subsection{Sampling}
\label{system:sampling}
Our system uses online sampling to meet interactive latency requirements for large datasets.
We support two sampling modes, simple random sampling and cluster sampling~\cite{lohr2009sampling}.
In simple random sampling, every tuple has an equal probability of being included in the sample. In the absence of indexes, this involves accessing every tuple of the dataset.
We use Bernoulli sampling semantics for simple random sampling.
In cluster sampling, different clusters are chosen randomly and all tuples within a cluster are included in the sample. 
This avoids accessing every tuple in the dataset.
The pros and cons of both sampling strategies are as follows.

\subsubsection{Execution Speed}

In function shipping, performing simple random sampling involves adding Bernoulli sampling-based scan operator and accessing the whole data set, while in cluster sampling, only a fraction of tuples are accessed.
In data shipping, to perform simple random sampling, the entire dataset needs to be transferred to the coordinator as the worker lacks computing resources required to perform sampling. 
The coordinator then samples the received data.
For cluster sampling, the coordinator only accesses a sample of the data, resulting in less network traffic.
Thus, cluster sampling is cheaper than simple random sampling for both shipping modes.

\subsubsection{Result Quality}
\label{system:result-quality}
Simple random sampling usually results in better sample quality than cluster sampling if the tuples are stored in non-random order.
A clustered index stores the data in a sorted order.
If the \texttt{GROUP BY} or \texttt{WHERE} clause contains any of the clustered index columns in order, cluster sampling can result in tuples and groups being respectively missed, causing sampling error to be large.

\section{Experiments}
\label{expt}

We extended our RDMA-aware query execution engine Pythia, a prototype open-source in-memory query
engine~\cite{liu2017design}, with sampling support.
We currently employ a single coordinator and a single worker setup.
They each have 512 GB of memory across two NUMA nodes, with each 
NUMA node having one Intel Xeon E5-2680v4 14-core processor.
They are connected by an EDR (100 Gb/s) InfiniBand network.

Our dataset has one table \texttt{R} with $10$ billion tuples, with each tuple having
two long integers \texttt{R.a} and \texttt{R.b} as attributes.
\texttt{R.a} is the primary key, which thereby ranges from 1 to the cardinality of the table,
and the distinct cardinality of \texttt{R.b} is varied.
We evaluate our system using the SQL query \texttt{SELECT R.b, COUNT(*) FROM R GROUP BY R.b}.
In the execution of the SQL query, records with the same value in \texttt{R.b} are aggregated
to a single record.
Hence the number of records in the result is the same as the number of distinct values of \texttt{R.b}
in the data.

\subsection{Changing Cardinality of Results}
In exploring the trade-offs between function and data shipping modes, a natural question to ask is, how does the size of the result affect their response times~(Section~\ref{system:function-vs-data-shipping})?

As the result size is non-deterministic with sampling, we turn off sampling in this experiment.
We vary the distinct cardinality of \texttt{R.b} from 1 thousand to 1 billion.
At the coordinator, we use all 28 cores for query execution, while the worker only uses 14 cores to simulate 
the additional workload in the worker node.
Figure~\ref{fig:varycardinality} shows that when the result size is less than or equal to 4 MB, function shipping has lower response time than data shipping.
This is because the size of the result which is transferred in function shipping is not large.
When the result size is equal to or larger than 8 MB, the saving in network traffic decreases
and function shipping has higher response time than data shipping.
Hence, data shipping is preferred when the result size is large.

\subsection{Changing Load on the Workers}
Another question is, how does the load on the worker and the choice of sampling method
affect the choice between function shipping and
data shipping?

We simulate different loads on the worker by varying the number of available cores from 1 to all
28 cores, and keeping the number of cores at the coordinator fixed at 28.
We set the distinct cardinality of \texttt{R.b} to be 2, and the query timeouts at 60 seconds.
The sampling rate is $10\%$ and we compare both cluster sampling and random sampling.
The result is shown in Figure~\ref{fig:varycores}.
The number of available worker cores has no impact on data shipping due to our use of RDMA.
The response time for function shipping decreases when the number of worker cores increases.
When the number of cores in the worker is 8 and 9, data shipping has higher performance for random
sampling but has lower performance for cluster sampling.
This is because random sampling is more computation
intensive and favors data shipping when the worker has limited computing resources.
For the same sampling method, we can see that data shipping has lower response time
when the number of cores is small and is up to 6.5$\times$ faster than function shipping,
as the saving in network traffic is offset by the slow workers in query execution.

How does the
performance change if we increase the size of the aggregation result?
As discussed in Section~\ref{system:function-vs-data-shipping}, the cost of function shipping
$\texttt{COST}(FS)$ increases when the result size increases.
Data shipping will be preferred when the result size increases and the cross point between
function shipping and data shipping will move to the right of the horizontal axis in Figure~\ref{fig:varycores}.

\subsection{Executing Multiple Queries}
Here, we look at the case where multiple queries are executed \emph{at the same time}.
How does executing multiple queries affect the decision to choose between function and data shipping in presence of sampling?

We use all 28 cores for both the worker and coordinator nodes.
The distinct cardinality of \texttt{R.b} is set to be 512 million
and set the sampling rate to $10\%$.
We run $Q1$ multiple times, ranging from 1 to 5.
Figure~\ref{fig:varyquery} shows that to achieve the identical sampling rate~($10\%$), random sampling is more expensive than cluster sampling.
Within the same sampling method, when running a single query, function shipping is better than data shipping.
This is because the size of the result transferred in function shipping is less than the size of the data transferred in data shipping.
However, data shipping becomes preferable over function shipping when the number of queries increases~($> 3$) in our setup.
This is due to result size increasing with the number of queries increasing, while the data transferred in data shipping stays the same.

\section{Related Work}
\label{rel-work}

RDMA has been studied in multiple database operations.
RDMA has been used to accelerate join execution.
Frey et al.~\cite{FreyGKTicdcs10} build a new join algorithm, \emph{cyclo-join}, which transfers data using RDMA.
Tinnefeld et al.~\cite{tinnefeldicde14} compare different join algorithms over RAMCloud, which is connected
with RDMA-enabled network.
Barthels et al.~\cite{BarthelsRackjoin15} study the radix join algorithm using RDMA to transfer data.
R{\"{o}}diger et al.~\cite{Rodiger2016} have designed \emph{flow-join}, which uses RDMA to deal with skew in join execution.
RDMA has also been used to accelerate data shuffling in parallel database systems.
R{\"{o}}diger et al.~\cite{Rodiger2015vldb} design a multiplexer which uses RDMA for data transfer.
As RDMA provides direct memory access to remote memory, M\"{u}hleisen et al.~\cite{Muhleisen2013} study the performance of accessing
remote memory in database systems; Li et al.~\cite{Li2016_Accelerating} use RDMA to directly access buffer pool in remote nodes.
RDMA is also studied in data processing.
Lu et al.~\cite{LuHadoopRdma13} accelerate Hadoop using RDMA.
Dragojevi\'{c} et al.~\cite{DragoFarm14} build a distributed computing platform, FaRM, with RDMA.
Wu et al.~\cite{gramsocc15} design a graph processing engine over FaRM.
Chen et al.~\cite{ChenRdmaEuro16} and Wei et al.~\cite{WeiRdmaSosp15} build distributed transaction processing systems using
RDMA.
Kalia et al.~\cite{KaliaOsdi16} implement RPC with RDMA and use RDMA-enabled RPC for distributed transaction processing.
RDMA is also used in key-value stores~\cite{Kaliasigcomm2014,Mitchellusenix2013}.

Sampling has been introduced in the context of databases by Olken et al~\cite{olken1993random}.
 The \emph{AQUA} system incorporated sampling into their real-world production environment, including supports for joins.
Different database systems such as \emph{SQL Server}, \emph{DB2}, \emph{AQUA}~\cite{acharya1999aqua}, \emph{Turbo-DBO}~\cite{dobra2009turbo}, \emph{BlinkDB}~\cite{agarwal2013blinkdb}, \emph{Quickr}~\cite{kandula2016quickr} have varying degrees of support for sampling.
Others have incorporated online sampling in the context of a session~\cite{kamat2014distributed, kamat2017session}.
 Sampling over joins has also received in-depth attention as sampling in many-to-many joins has theoretical and practical constraints~\cite{chaudhuri1999random, kamat2017unified}.
Online aggregation~\cite{hellerstein1997online} introduced the notion of decreasing error \emph{during} partial query execution.
 Researchers have also looked at using data cubes and binning to provide scalable interactive visualizations~\cite{lins2013nanocubes, kamat2017infiniviz}.
In contrast, our system is the first to consider the implications of using sampling in the context of RDMA.

\section{Conclusions and future work}
\label{conclusion}

In this paper, we compare how RDMA and fast networks affect query execution strategies for
interactive queries with sampling.
While function shipping was the norm,
interactive big data analytics should take the amount of data transferred, the CPU utilization, the sampling methods and the number of queries executed on the data set into account when choosing query execution strategies.
Looking ahead, one possible direction is to build a cost model which takes these factors into account to predict the cost of different execution strategies and to pick the optimal execution strategy.

\vspace{1em}
\section*{Acknowledgements}
This material is based upon work supported by the National Science Foundation
under grants IIS-1422977, IIS-1527779, CAREER IIS-1453582,     
CCF-1816577 and CNS-1513120.
Any opinions, findings, and conclusions or recommendations expressed in this
material are those of the author(s) and do not necessarily reflect the views of
the National Science Foundation.

\small
\bibliographystyle{abbrv}
\bibliography{allfilesInOne}

\end{document}